\documentclass[aps,pre,preprint,groupedaddress,showpacs]{revtex4}

\usepackage{graphicx}
\usepackage{dcolumn}
\usepackage{bm}

\begin{document}


\title{Model for the spatio-temporal intermittency of the energy dissipation
in turbulent flows}


\author{Fabio Lepreti} 
\email[Corresponding author. Email address: ]{lepreti@fis.unical.it}
\author{Vincenzo Carbone}
\author{Pierluigi Veltri}
\affiliation{Dipartimento di Fisica, Universit\`a della Calabria,
Via P. Bucci 31/C, I-87036 Rende (CS), Italy}
\affiliation{Consorzio Nazionale Interuniversitario per le Scienze
Fisiche della Materia (CNISM),~Unit\`a~di~Cosenza, Italy}


\date{\today}

\begin{abstract}
Modeling the intermittent behavior of turbulent energy dissipation
processes both in space and time is often a relevant problem when
dealing with phenomena occurring in high Reynolds number flows,
especially in astrophysical and space fluids.
In this paper, a dynamical model is proposed to describe the 
spatio-temporal intermittency of energy dissipation rate in a turbulent
system.
This is done by using a shell model to simulate the turbulent cascade
and introducing some heuristic rules, partly inspired by the
well known $p$-model, to construct a spatial structure of the energy
dissipation rate.
In order to validate the model and to study its spatially intermittency
properties, a series of numerical simulations have been performed. 
These show that the
level of spatial intermittency of the system
can be simply tuned by varying a single parameter of the model
and that scaling laws in agreement with those obtained from
experiments on fully turbulent hydrodynamic flows can be recovered.
It is finally suggested that the model could represent a useful tool
to simulate the spatio-temporal
intermittency of turbulent energy dissipation in those high Reynolds
number astrophysical fluids where impulsive energy release processes
can be associated to the dynamics of the turbulent cascade.
\end{abstract}


\maketitle



\section{Introduction}
The dynamics of fluids and plasmas, both in laboratory experiments
and in astrophysical or geophysical systems, is very often
characterized by the presence of turbulent motions~\cite{frisch,choudhuri}.
In several contexts of astrophysics and space physics, it is extremely 
important to model some of the effects related
to the turbulent dynamics. In particular, describing in a proper way
the spatio-temporal intermittency of the turbulent energy dissipation
process is one of the basic ingredients for the study of several
astrophysical systems. As relevant examples, we can
consider the active regions of the solar corona \cite{dmitruk97,veltri05},
the interstellar medium \cite{franco99},
and accretion disks \cite{balbus98}.

Intermittency is one of the most investigated problems
in the field of fully developed turbulence (see \cite{frisch} and
references therein). Among the many approaches used for the study
of intermittency in turbulence, here we want to briefly recall only some
of them, which are related to the work presented in this paper.
A number of random cascade models
(see {\it e.g.}~\cite{k62,frisch78,benzi84,meneveau87})
were initially proposed to reproduce the observed intermittency corrections 
(see {\it e.g.}~\cite{anselmet84}) to the scaling
laws of the classical Kolmogorov theory of turbulence~\cite{k41}.
Another interesting approach to the modeling of intermittency
of the turbulent energy cascade is based on the use of dynamical
deterministic models known as shell models (see the reviews by
\citet{bohr}, \citet{giuliani99}, \citet{biferale03}).
More recently, several relevant developments have led to
a beginning of a deeper understanding of the intermittency phenomenon.
To mention but a few: the role of Lagrangian conservation laws
\cite{falkovich01} and nonlocal interactions \cite{laval01} on
intermittency, and the introduction of new multifractal
approaches for the description of velocity increments
statistics \cite{chevillard06}.

Besides these important theoretical advances, there are several more specific
situations where a simple dynamical system modeling of the intermittency  in the
turbulent cascade can be extremely helpful.
This can be the case of astrophysical and space fluids,
where, due to the extremely large Reynolds numbers, dynamical
models which are able to simulate the turbulent cascade and the
related energy dissipation processes in Reynolds number regimes which are
not far from the real ones (at least with respect to direct numerical simulations)
can represent an essential ingredient for the modeling of such physical systems.
An example is given by the recent
applications of shell models to the description of the
statistical properties of solar flares~\cite{b99,lepreti04},
and to the nanoflares occurring in solar coronal loops~\cite{nigro04}.
In this framework, it is worth to point out that shell models provide
only a temporal description of the intermittency properties since they lack
any spatial information. The possibility to have a dynamical ``shell-like''
model capable of reproducing some intermittency properties both in space
and time would thus be attractive.

For the reasons explained above, in this work we propose a simple method 
to model the intermittent character of energy dissipation in a turbulent
system in both space and time,
by using a shell model together with some rules inspired to some extent
by the well known turbulence $p$-model \cite{meneveau87}. The paper is
organized as follows. In Section
\ref{sec-background} we recall the main ideas concerning shell models
and the $p$-model, in Section \ref{sec-method} we give a description of 
the proposed method, in Section \ref{sec-procedure} we provide some details
about the numerical procedure, in Section
\ref{sec-result} we show the results of the spatially intermittency analysis
performed in order to validate the proposed model,
while the conclusions are drawn in Section \ref{sec-conclu}.


\section{Background}
\label{sec-background}

\subsection{Shell model}
Shell models were introduced in 70's by~\citet{obukhov71},~\citet{gledzer73}, 
and~\citet{desnyansky74} in the context of hydrodynamic turbulence
and, since then, used extensively both in hydrodynamics (see e.g.
\cite{yamada87,jensen91}) and magnetohydrodynamics (MHD) (see e.g.
\cite{gloaguen85,biskamp94,frick98,giuliani98}).
They are based on a set of coupled nonlinear ordinary differential 
equations which describe the dynamics of the turbulent energy cascade 
in the wave-vector space. The dynamical and statistical behavior 
of shell models have been investigated in detail in many works
(see \cite{bohr,giuliani99,biferale03} and references therein)
and it has been shown that they are able to describe several
properties of the turbulent energy cascade process. The main advantage
of shell models is that they can be investigated
through numerical simulations at high Reynolds numbers much more
easily than Navier-Stokes (N-S) or MHD equations, due to the reduced 
number of degrees of freedom. On the other hand, an obvious minus of
these models is the absence of any information about spatial structures.

Shell models are built up by dividing the wavevector space in a 
discrete number of shells of radius $k_n = k_0 \lambda^n$, with
$\lambda > 1$ fixing the shell logarithmic spacing (usually
$\lambda = 2)$ and $n=1,2,...,N$. Each shell is associated with
a dynamical complex variable $u_n(t)$ which represents the time
evolution of velocity fluctuations at scale $\ell_n \sim k_n^{-1}$.
The evolution equations for the variables $u_n(t)$ are written
by introducing nonlinear terms in the form of quadratic couplings
between neighbouring shells. The coupling coefficients are chosen
to satisfy scale invariance and the conservation of the ideal 
invariants, as for example the total energy and the kinetic helicity for
N-S equations.

For this work we use the hydrodynamic shell model proposed by
\citet{lvov98}, also known as Sabra model. The evolution equations
of the shell variables are
%
%
\begin{eqnarray}
{d u_n \over dt} = & i k_n \left( u_{n+2} u_{n+1}^*
- {1 \over 4} u_{n+1} u_{n-1}^* 
+ {1 \over 8} u_{n-1} u_{n-2} \right) & \nonumber \\ 
& -\nu k_n^2 u_n + f_n \; , &
\label{eq-shell}
\end{eqnarray}
where $\nu$ represents the kinematic viscosity, and $f_n$ is a
forcing term usually acting on some low wavenumber shells.
From Eq. (\ref{eq-shell}), we can derive the evolution equation
for the $n$-th shell kinetic energy $E_n = (u_n u_n^*)/2$:
\begin{equation}
{d E_n \over dt} = - \Pi_n -2 \nu k_n^2 E_n + \Re (u_n^* f_n) \; ,
\end{equation}
where
%
%
\begin{eqnarray}
& \Pi_n =
k_n \Im ( u_n^* u_{n+2} u_{n+1}^*
+ {1 \over 4} u_n u_{n+1}^* u_{n-1} &\nonumber \\
& + {1 \over 8} u_n^* u_{n-1} u_{n-2} ) &
\end{eqnarray}
gives the kinetic energy flux through the $n$-th shell (the symbols
$\Re$ and  $\Im$ denote the real and imaginary parts respectively
of a complex number).

The total energy dissipation rate $\varepsilon(t)$ can be defined as
\begin{equation}
\varepsilon(t) = \nu \sum_{n=1}^N k_n^2 |u_n|^2 \; .
\end{equation}

\subsection{The $p$-model}
The $p$-model has been designed to describe the observed multifractal
behavior of the energy dissipation rate in fully turbulent flows  
\cite{meneveau87}. Without loss of generality, in this paper we consider, 
for simplicity, a one-dimensional
spatial domain. In this case, the total dissipation rate $\varepsilon_r(x)$
in the segment [$x$,$x+r$]
is equal to $\varepsilon_r(x) = \int_x^{x+r} \varepsilon(x) dx$,
$\varepsilon(x)$ being the energy dissipation rate in the $x$ position.
In the $p$-model, an interval of size $r$ breaks down into two
subintervals of size $r/2$, and the energy flux to these smaller eddies
proceeds unequally. A fraction $p$ (with $0.5 \le p \le 1$) of the
dissipation contained in the parent interval is distributed equally
on one of the two subintervals (left or right with equal probability),
and the remaining $(1-p)$ fraction on the other subinterval. This 
process starts from the integral scale $L$ (where we have only one
interval) and is repeted until segments of size $\eta$ (corresponding
to the dissipative scale) are created. It has been shown in Ref.
\cite{meneveau87} that using $p = 0.7$ the multifractal spectrum
of the synthetic dissipation signal obtained through the $p$-model
reproduces extremely well the results of experiments. 

\section{The model}
\label{sec-method}

In a few words, the basic idea of the method proposed here
consists in using a shell model to describe the dynamics of the
turbulent cascade process and in providing a spatial structure to
the energy dissipation using some rules, which partly recall the 
$p$-model, to distribute in space the energy fluxes given by the 
shell model.

We consider a one-dimensional spatial domain whose size is denoted
by $L$. As a base for the construction of the spatial energy structure
we use a hierarchy of $N$ scales $\ell_n= 2^{1-n} L$ ($n=1,2,...,N$). 
For each scale $n$ we can define a set of
$2^{n-1}$ disjoint segments of size $\ell_n$ which cover the
spatial domain. Let us note that this is the same hierarchical 
structure as in the $p$-model, that is, each segment at the scale
$n$ can be considered as parent of two corresponding segments
at the scale $n+1$ which have half the size and cover the same
subinterval as the parent.

Let us now suppose that the $N$ scales of this hierarchy are
associated with  the $N$ shells of the shell model, and that
$\ell_n = 1/k_n$.
At each time step $t_i$ (i=1,2,...) of the numerical solution
of the shell model
equations, we can calculate the energy increment $\Delta E_n(t_i)$ of 
the $n$-th shell as
\begin{equation}
\Delta E_n(t_i) = E_n(t_i) - E_n(t_{i-1}) \; .
\end{equation}
These increments are used to construct a spatial energy structure
which evolves in time parallely to the shell model as explained
below.

The increments $\Delta E_n(t_i)$ are distributed over the spatial 
grid of the corresponding scale using the following criteria:

\begin{itemize}

\item
If $\Delta E_n(t_i) > 0$ we first divide the energy $\Delta E_n(t_i)$
among the segments at the scale $n-1$ (which thus play the role of
parent segments) proportionally to the energy
contained in them at the scale $n-1$.
The energy increment thus obtained for each parent segment
is then transferred to the corresponding two daughter segments
in the same way as in the $p$-model, that is, adding a fraction
$p$ of the increment ($0.5 \le p \le 1$) to one of the daughters 
(left or right with
equal probability) and the remaining $(1-p)$ fraction to the other
daughter.

\item
If $\Delta E_n(t_i) < 0$ the increment is subtracted from the energy
at the scale $n$ in such a way that each segment is depleted by
a fraction of $\Delta E_n(t_i)$ proportional to the energy content
of the segment itself at the same scale.

\end{itemize}

As a result of the procedure described above,
to each one of the $2^{n-1}$ segments which cover the domain at the
scale $n$ is attributed a kinetic energy $E_l^{(n)}(t)$
(where $l=1,...,2^{n-1}$ is the index denoting the segments at the
scale $n$). The total energy at the scale $n$ equals the kinetic
energy of the corresponding shell, that is, 
$\sum_{l=1}^{2^{n-1}} E_l^{(n)}(t) = |u_n(t)|^2/2$.

In order to have an evolution of the spatial energy structure, the
spatial distribution of the $p$ and $(1-p)$ values is changed
during the time evolution. Two different methods, described below,
were used to 
perform this changes in time and we denote by Model A and Model B
the two versions of the model corresponding respectively
to these two methods. 

{\em Model A}. 
In the first version of the model the changes of the probabilities
in space are done at the same time instant for all the segments
at a given scale $n$. Let us suppose
that the last change in the $p$ distribution at the scale $n$
occurs at the time step $t_j^{(n)}$. At each time step $t_i$
we calculate the {\em instantaneous eddy turnover time}
$\tau_e^{(n-1)} (t_i) = [k_{n-1} u_{n-1}(t_i)]^{-1}$ for the scale 
$n-1$ and compare it to the time elapsed from the last change 
$\Delta t^{(n)} = t_i -t_j^{(n)} $.
If $\Delta t^{(n)} > \tau_e^{(n-1)}$, the $p$ spatial distribution
at the scale $n$ is redrawn. This procedure is followed for each 
$n>1$.

{\em Model B}.
In the second version the spatial changes of the probabilities
are done independently for the different segments at a given
scale $n$. Indeed one of such changes always involves a couple of
segments $l$ and $l+1$ (where $l$ is an odd integer number), because
if the probability is $p$ for the segment $l$ it must be $1-p$ for
the segment $l+1$ and viceversa. As a consequence, if we denote by 
$t_j^{(n,l)}$ the last change for the
segment $l$ at the scale $n$, we have that $t_j^{(n,l)}=t_j^{(n,l+1)}$
for a couple of neighbouring segments with $l$ odd integer. 
At each time step $t_i$, we compute
the {\em local instantaneous eddy turnover time} for the
corresponding {\em father segment} at the scale $n-1$
(whose index is $l_f = {l+1 \over 2}$), that is,
$\tau_e^{(n-1,l_f)} (t_i) = 
[k_{n-1} \sqrt{2 E_{l_f}^{(n-1)} (t_i)}]^{-1}$
and compare it to the time elapsed from the last change 
$\Delta t^{(n,l)} = \Delta t^{(n,l+1)} = t_i -t_j^{(n,l)}$.
If $\Delta t^{(n,l)} > \tau_e^{(n-1,l_f)} (t_i)$ 
the probability values are redrawn for the segments $l$ and $l+1$. 
This procedure is followed for each $n>1$.

The two procedures described above aim to describe phenomenologically 
the correlations arising in the cascade due to the scaling of the 
eddy turnover times. The main difference between them consists
in the fact that in Model B we try to take into account also
the local dynamics of the turbulent cascade on the spatial domain.

In both the versions of the model, summing the contribution coming 
from all the scales, we obtain
the spatial shape of the energy density $w(x,t)$ as
\begin{equation}
w(x,t) = \sum_{n=1}^N {E_{l_n(x)}^{(n)} \over \ell_n} \; ,
\end{equation}
where $x$ denotes the grid position, corresponding to the smallest
scale grid spacing, and  $l_n(x) = \mathrm{Int}[(x-1) / \ell_n] + 1$
(Int denotes the integer part of a real number).
We can now define also an energy dissipation rate which depends 
on the spatial coordinate as
\begin{equation}
\varepsilon(x,t) = \nu \sum_{n=1}^N k_n^2 {E_{l_n(x)}^{(n)} \over \ell_n}
\; .
\label{eq-diss-xt}
\end{equation}
%


\section{Numerical procedure}
\label{sec-procedure}

\begin{figure}
\includegraphics[width=\columnwidth]{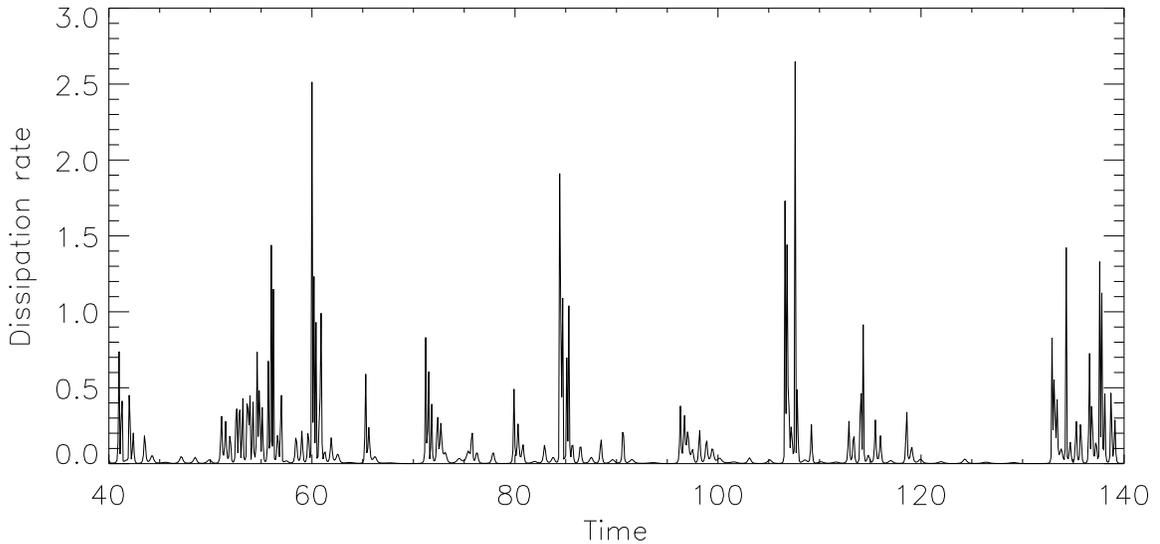}
\caption{A sample of the total energy dissipation rate given by the 
shell model.}
\label{fig-diss-shell}
\end{figure}

The shell model equations Eqs. (\ref{eq-shell}) have been numerically 
solved 
using a 4-th order Runge-Kutta scheme. The parameters used in the
simulations are $N=15$, $\nu=10^{-4}$, $k_0=2^{-5}$, and $\lambda=2$. 
We used an external forcing term applied on the 3-rd and 4-th shell given
by $f_3 = f_4 = 0.1(1+i)$. With these parameters the Reynolds
number is $Re \simeq 10^5$ and the large scale eddy turnover time 
$\tau_e \simeq 40$.
A sample of the total energy dissipation rate $\varepsilon(t)$ given
by the shell model is shown in Fig. \ref{fig-diss-shell}.

The number of grid points in the 1-D spatial domain is $2^{14}$.
Another free parameter of the model is the $p$ value used
to construct the spatial structure of the energy dissipation.
It should be clear to the reader that it is not necessary to
use the value $p=0.7$ which allows the $p$-model to reproduce
the multifractal structure of the energy dissipation rate
observed in experiments of fully developed turbulence \cite{meneveau87}. 
This is why the models proposed here, although inspired to some
extent by the $p$-model, are substantially different from it,
being characterized also by the dynamics provided by the shell model
and by the evolution of the $p$ spatial distribution.
The multifractal properties of the spatial energy dissipation
given by the model change in time. More precise indications
about the values to be attributed to the $p$ parameter can be 
found from the application of the model to well defined physical
situations, a question which we plan to further investigate in the future.
For this work, we have performed numerical simulations
using the values of $p = 0.7, 0.8, 0.9, 1$ for Model A,
and $p=0.7, 0.8$ for Model B.
We would like to remark that changing the spatial
distribution of the $p$ and $1-p$ values according to the
instantaneous eddy turnover times at each scale, as described in
the previous section, is necessary for describing the intermittent
behavior of the energy dissipation. As a confirmation, we performed
a simple test in which we modified the spatial distribution of the
probabilities at each time step and obtained that in this case the
energy dissipation rate is not intermittent for any value of $p$.

\section{Analysis of the spatial intermittency}
\label{sec-result}

In order to validate the proposed models,
we performed an analysis of the spatial intermittency properties of 
the energy dissipation rate for different values of the $p$ parameter. 
This has been done both by calculating the kurtosis of $\varepsilon(x)$
and by looking for the presence of 
multifractal scaling laws in $\varepsilon(x)$
\cite{parisi85,paladin87,sreenivasan91}. In this way, 
some comparisons to the scaling exponents
of structure functions found in turbulence experiments can
also be made.

The starting point for multifractal analysis is the definition
of the probability measure
\begin{equation}
\mu_i(r) = {\chi_i(r) \over \chi(L)} \; ,
\end{equation}
where
\begin{equation}
\chi_i(r) = \int_{S_i(r)} \varepsilon(x) dx \; .
\end{equation}
$S_i(r)$ represents a hierarchy of disjoint segments of size $r$
covering the domain $L$. We can calculate the so called generalized
dimensions $D_q$ \cite{hentschel83} by looking at the scalings of 
the $q$-th order moments of $\mu_i(r)$ vs. $r$:
\begin{equation} 
\langle \mu^q \rangle = \sum_i \mu_i^q(r) \sim r^{(q-1)D_q} \; .
\label{eq-moments}
\end{equation}
The largest values of $q$ amplify the contribution given to 
$\langle \mu^q \rangle$ by the most intermittent regions of the measure,
while for small values of $q$ the major contribution is due to the most 
regular regions. If the probability measure is globally self-similar
(i.e. non intermittent),
$D_q$ is constant and it corresponds to the fractal dimension of the 
measure. Conversely, if $D_q$ is not constant, the scaling laws are
said to be anomalous and the measure can be described as a multifractal
object. In this case, it can also be shown that $D_q$ is a nonincreasing
function of $q$ \cite{paladin87}.

The generalized dimensions $D_q$ can also be related to the scaling
exponents $\zeta_q$ of the velocity structure functions, which are
measured in fluid flows and represent the benchmark for the nonlinear
energy cascade modeling. These exponents are defined by
\begin{equation}
\langle \delta v_r^q \rangle =
\langle \left[ v(x+r) - v(x) \right]^q \rangle
\sim r^{\zeta_q} \; .
\end{equation}
It can be shown \cite{meneveau87a} that
\begin{equation}
\zeta_q = {q \over 3} + \left( {q \over 3} -1 \right) 
\left( D_{q/3} - d \right) \; ,
\label{eq-zetaq-dq}
\end{equation}
where $d$ represents the topological dimension of the support,
in our case $d=1$.

We show in the next two subsection the results obtained for
Model A and Model B respectively.

\subsection{Model A}
\begin{figure}
\includegraphics[scale=0.8]{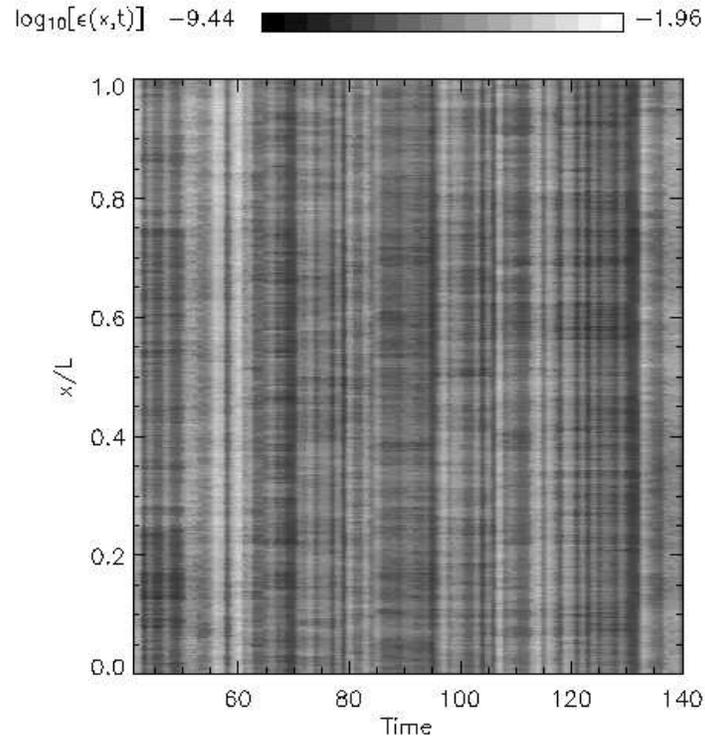}
\includegraphics[scale=0.8]{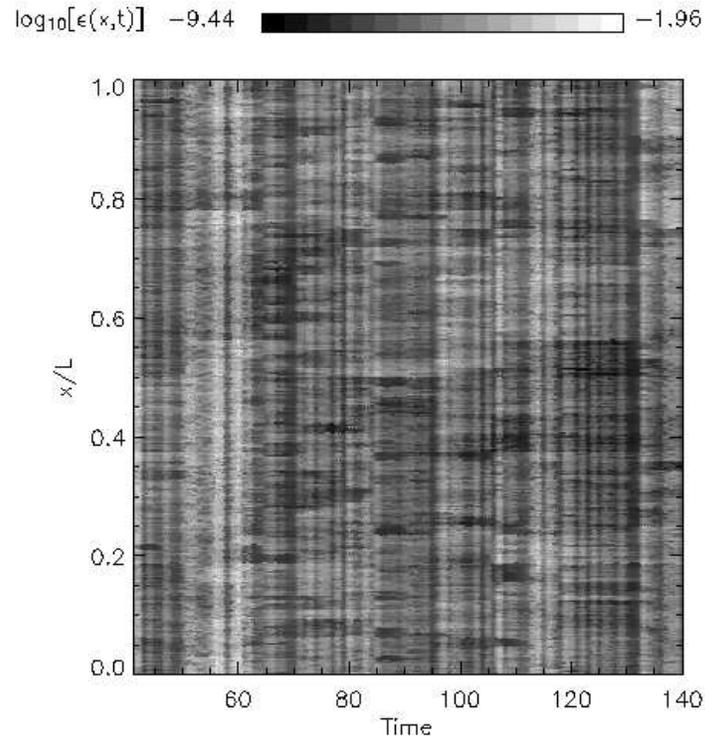}
\caption{Space-time structure of the energy dissipation rate $\varepsilon(x,t)$
in Model A for $p=0.8$ (top panel) and $p=1$ (bottom panel). The grey levels 
refer  to the logarithm of $\varepsilon(x,t)$.}
\label{fig-diss}
\end{figure}

In Fig \ref{fig-diss}
the space-time structure of the energy dissipation rate $\varepsilon(x,t)$,
calculated according to Eq. (\ref{eq-diss-xt}),
is shown for $p=0.8$ and $p=1$. 
For the sake of clarity the grey levels refer to the
logarithm of $\varepsilon(x,t)$.  It can be seen that $\varepsilon(x,t)$
becomes more and more fragmented in space as $p$ increases,
as one would expect.
%
%
%
%

\begin{figure}
\includegraphics[scale=0.65]{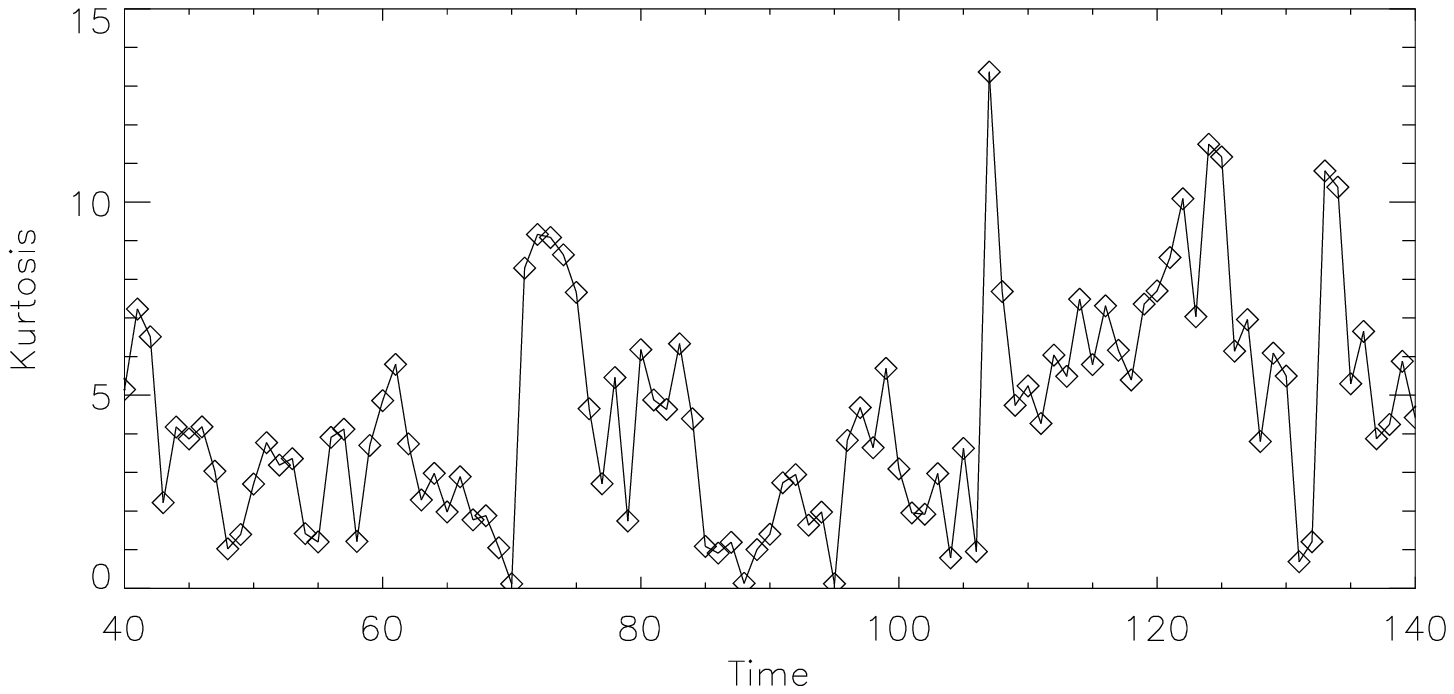}
\includegraphics[scale=0.65]{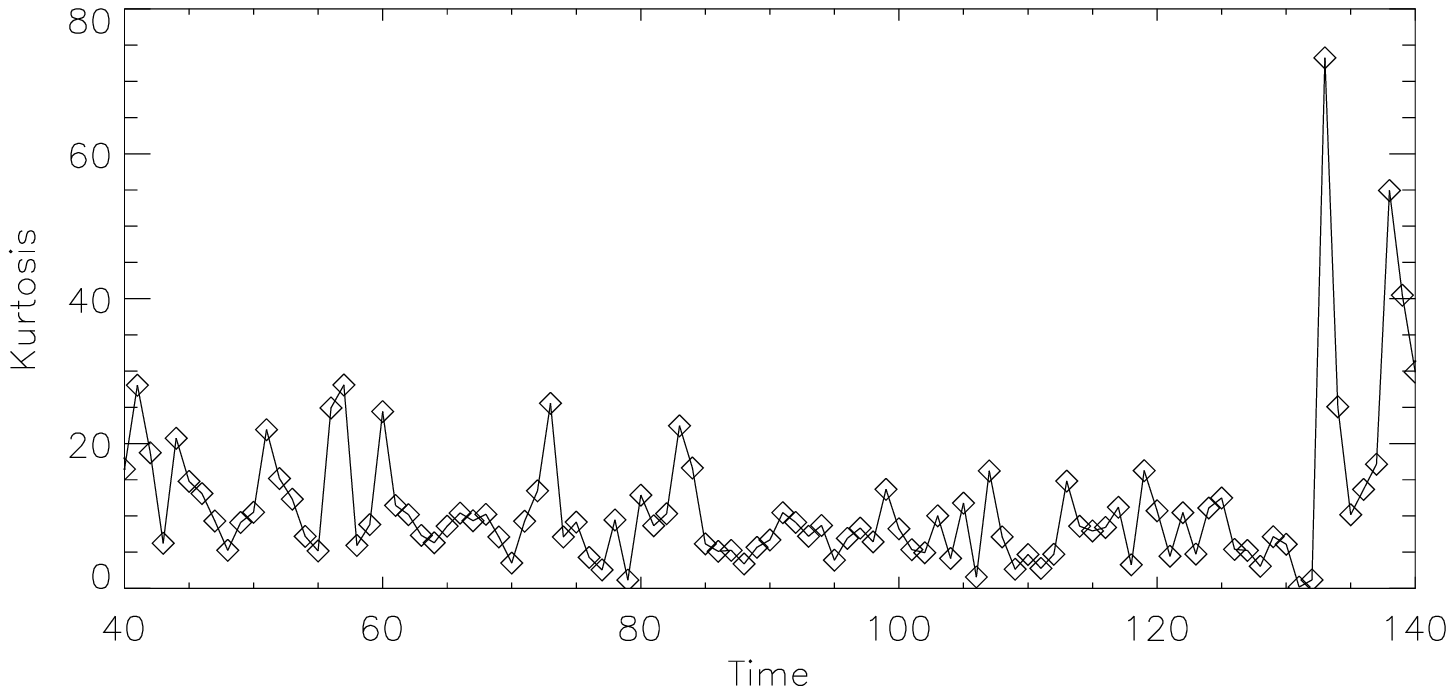}
\includegraphics[scale=0.65]{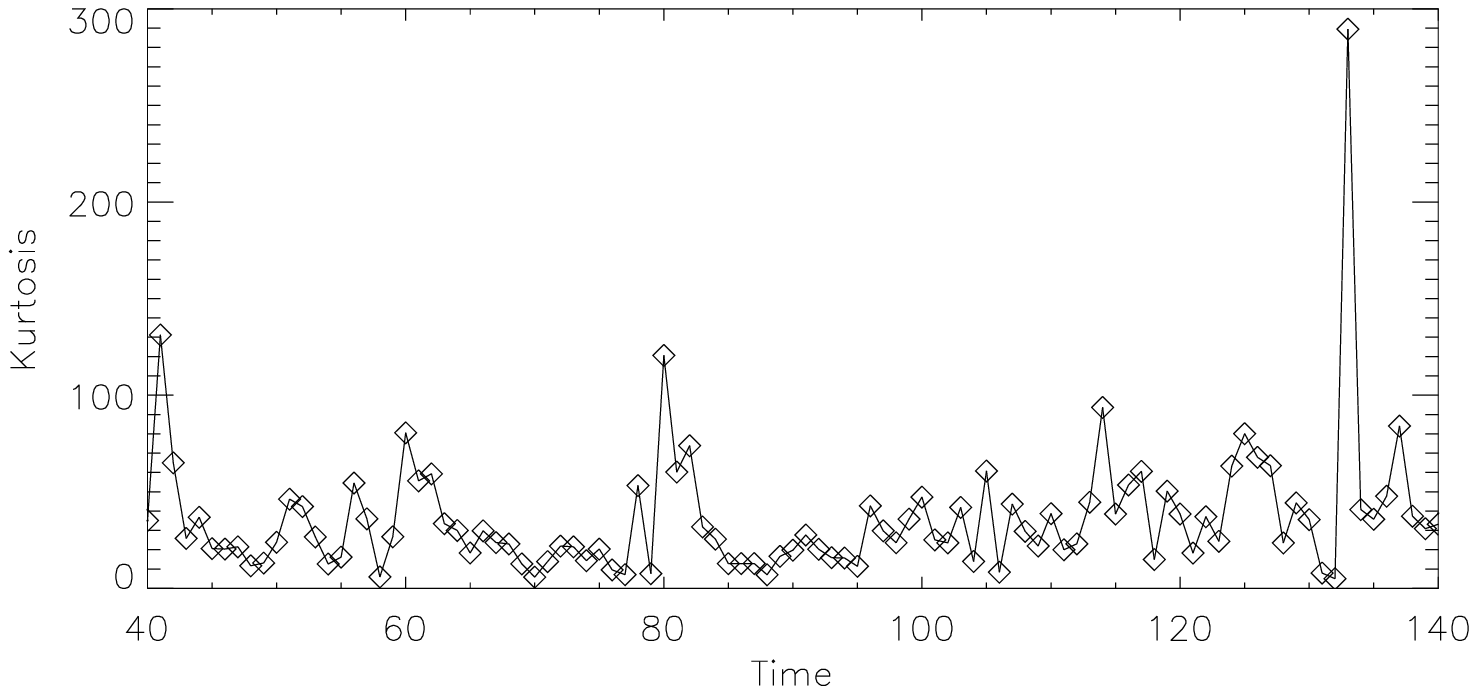}
\includegraphics[scale=0.65]{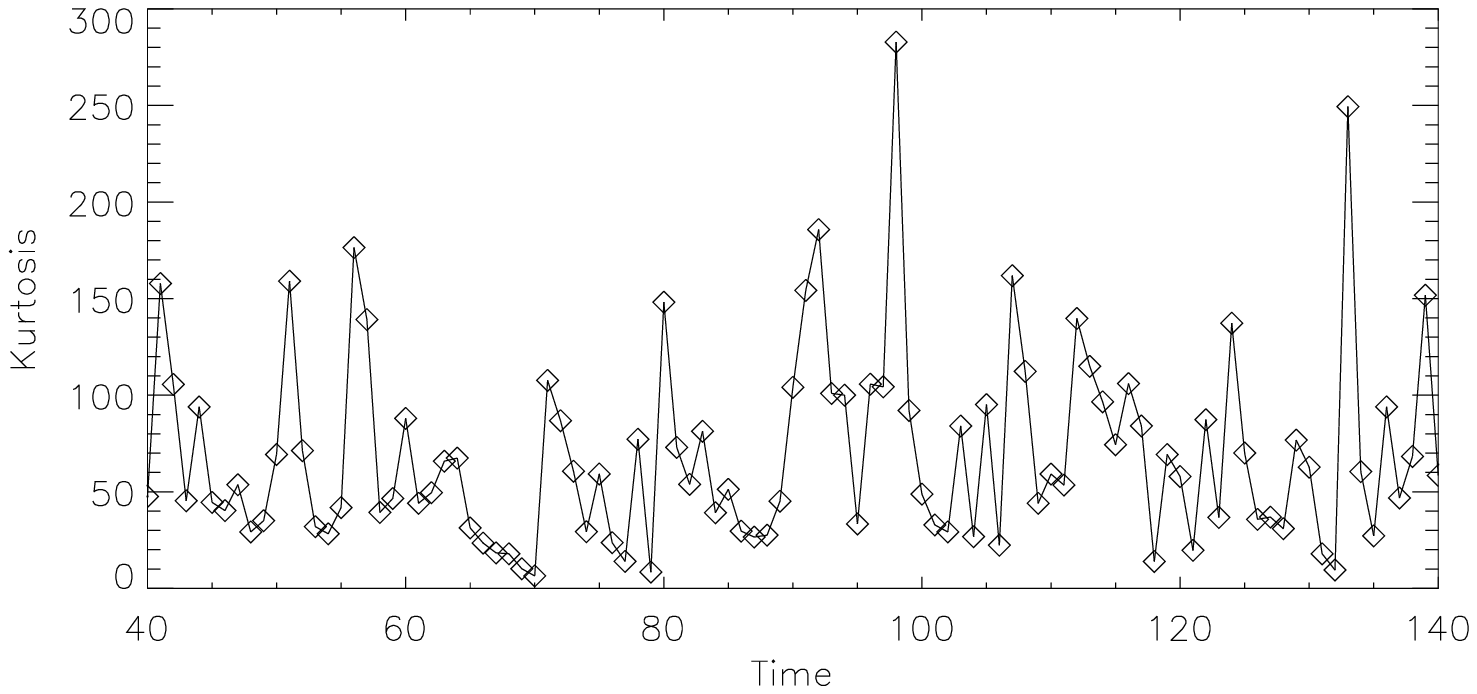}
\caption{Time evolution of the kurtosis of the spatial energy 
dissipation rate $\varepsilon(x)$ in Model A for the four different 
values of $p$ used: from top to bottom
 $p=0.7$, $p=0.8$, $p=0.9$, and $p=1$ respectively.}
\label{fig-kurto}
\end{figure}

\begin{figure}
\includegraphics[scale=0.65]{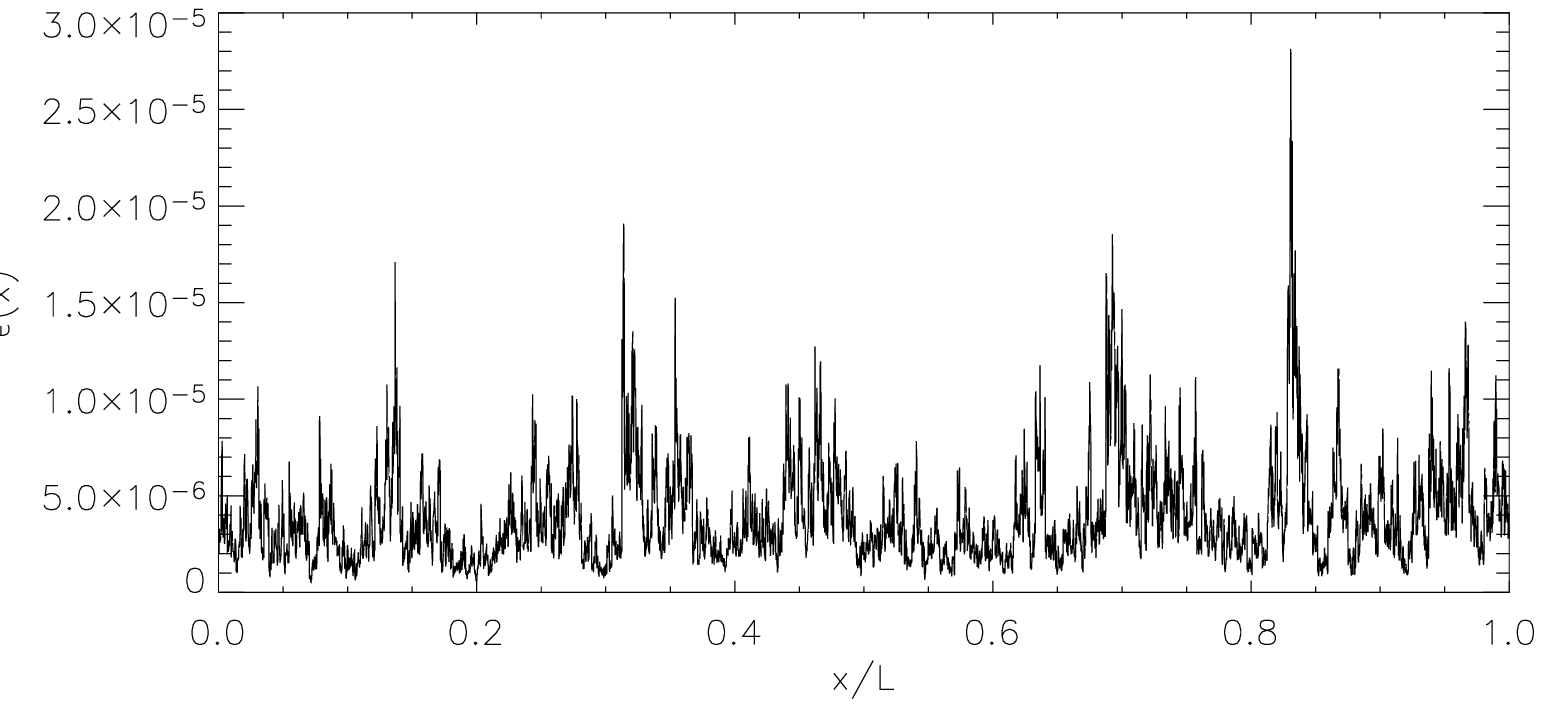}
\includegraphics[scale=0.65]{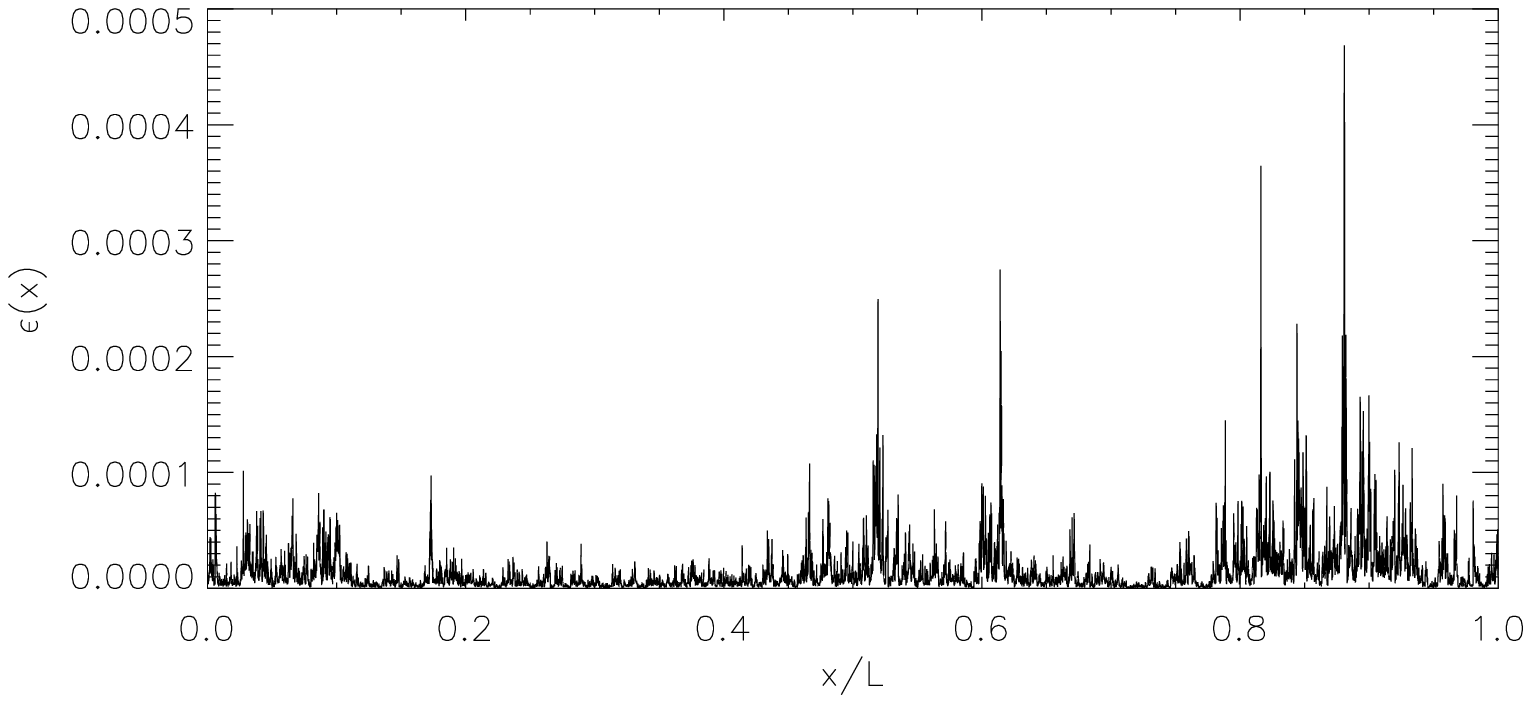}
\includegraphics[scale=0.65]{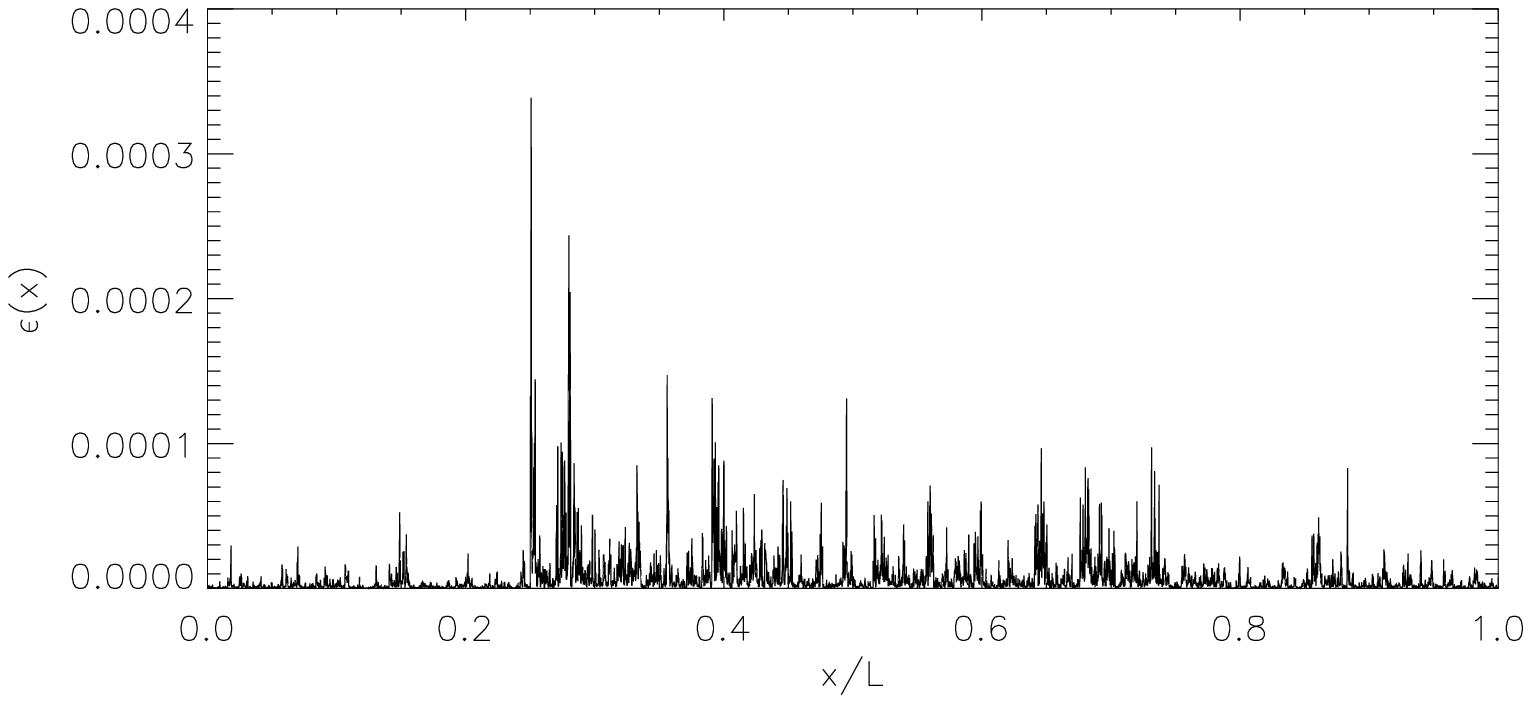}
\includegraphics[scale=0.65]{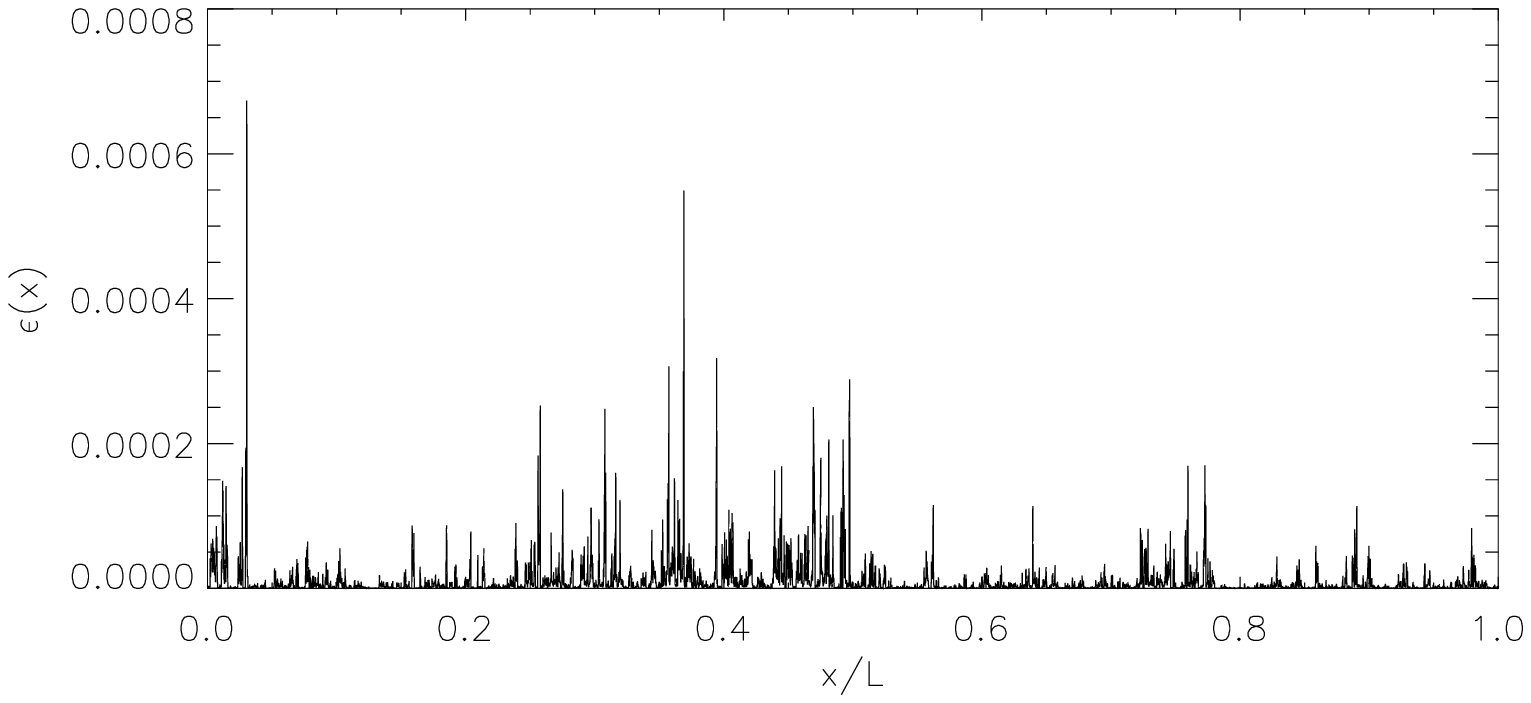}
\caption{Energy dissipation rate on the spatial domain in Model A: 
from top to bottom,
$p=0.7$ and $t=73$, $p=0.8$ and $t=133$, $p=0.9$ and $t=80$, $p=1$ and
$t=51$.}
\label{fig-ediss}
\end{figure}

To give a first indication on the intermittency properties of the
spatial energy dissipation $\varepsilon(x)$, the time evolution of the
kurtosis of $\varepsilon(x)$ for the four different values of $p$ used
is shown in Fig. \ref{fig-kurto}.
The increase of the typical values of the kurtosis confirms
that the level of intermittency is significantly enhanced
as $p$ increases from 0.7 up to 1. In Fig. \ref{fig-ediss}
the spatial structure of the energy dissipation rate $\varepsilon(x)$
is shown for four fixed time instants (one for each
of the different $p$ values used) where the kurtosis shown in
Fig. \ref{fig-kurto} displays a peak. The four time instants
chosen are $t=73$ for $p=0.7$, $t=133$ for $p=0.8$, $t=80$
for $p=0.9$, and $t=51$ for $p=1$. 
This figure shows that at some positions the quantity $\varepsilon(x)$
shows very strong bursts, which appear to become stronger and stronger
as $p$ increases.

\begin{figure}
\includegraphics[width=\columnwidth]{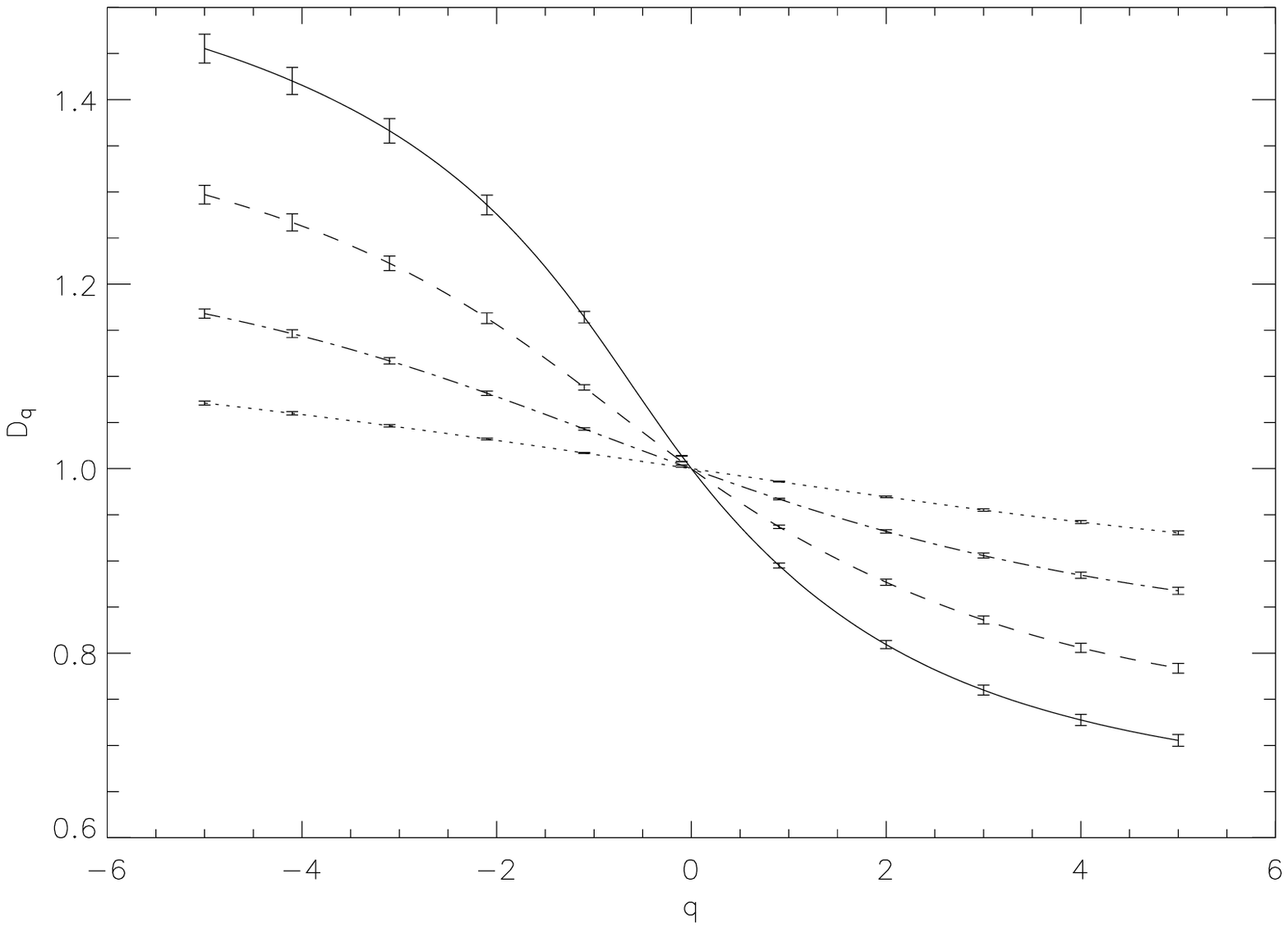}
\caption{Generalized dimensions $D_q$ of the spatial energy 
dissipation rates in Model A for $p=0.7$ ({\em Dotted line}),
$p=0.8$ ({\em Dash-Dotted}), $p=0.9$ ({\em Dashed line}),
and  $p=1.0$ ({\em Solid line}). The $D_q$ values were calculated
as time averages over all the time instants considered (see text).
Error bars, representing standard
deviation errors, are also reported for some values of $q$.}
\label{fig-dq}
\end{figure}

The multifractal analysis was performed by calculating the moments 
$\langle \mu^q \rangle$ given in Eq. (\ref{eq-moments}) for $-5 \le q \le 5$
at 100 different time instants, namely $t=41,42,...,140$.
A good scaling region extending almost over the whole $r$ range was
found for all the $q$'s. The generalized dimensions $D_q$ were calculated 
as averages over all the time instants considered. The plot
of $D_q$ vs. $q$ obtained for the four values of $p$ used is shown 
in Fig. \ref{fig-dq}.
It can be seen that the spectrum of $D_q$ values becomes wider as $p$
increases. This is a consequence of the enhanced level of intermittency
which produces stronger and more localized dissipation bursts for
larger values of $p$.

The scaling exponents $\zeta_q$ as estimated from the energy 
dissipation scalings in Model A for the four values 
of $p$ considered are reported in Table \ref{tab-vscal}.
%
  \begin{table*}
  \caption{Scaling exponents $\zeta_q$ for the velocity
  structure functions as 
  estimated from the energy dissipation scalings in Model A
  for the four values of $p$ considered. In the last column we report
  the velocity structure functions exponents computed from a wind tunnel 
  experiment \cite{ruiz95}.}   
  \begin{center} 
  \begin{tabular}{cccccc} 
  \hline
  $q$ \hspace{0.7cm} & $p$=0.7 \hspace{0.7cm} & $p$=0.8 \hspace{0.7cm} & 
  $p$=0.9 \hspace{0.7cm} & $p$=1.0 \hspace{0.7cm} & wind tunnel\\ 
  \hline
  1 \hspace{0.7cm} & $0.337 \pm 0.001$ \hspace{0.7cm} & 
  $0.342 \pm 0.002$ \hspace{0.7cm} & $0.350 \pm 0.005$ \hspace{0.7cm} &
  $0.362 \pm 0.008$ \hspace{0.7cm} & $0.37 \pm 0.01$ \\
  2 \hspace{0.7cm} & $0.670 \pm 0.001$ \hspace{0.7cm} & 
  $0.675 \pm 0.002$ \hspace{0.7cm} & $0.683 \pm 0.005$ \hspace{0.7cm} &
  $0.694 \pm 0.007$ \hspace{0.7cm} & $0.70 \pm 0.01$ \\
  3 \hspace{0.7cm} & $1.00$ \hspace{0.7cm} & $1.00$ \hspace{0.7cm} &
  $1.00$ \hspace{0.7cm} & $1.00$ \hspace{0.7cm} & $1.00$ \\
  4 \hspace{0.7cm} & $1.326 \pm 0.002$ \hspace{0.7cm} &
  $1.317 \pm 0.004$ \hspace{0.7cm} & $1.304 \pm 0.008$ \hspace{0.7cm} &
  $1.285 \pm 0.012$ \hspace{0.7cm} & $1.28 \pm 0.02$ \\
  5 \hspace{0.7cm} & $1.650 \pm 0.005$ \hspace{0.7cm} &
  $1.628 \pm 0.011$ \hspace{0.7cm} & $1.596 \pm 0.020$ \hspace{0.7cm} &
  $1.554 \pm 0.027$ \hspace{0.7cm} & $1.54 \pm 0.03$ \\
  \hline
  \end{tabular} 
  \label{tab-vscal} 
  \end{center} 
  \end{table*}
As a comparison, the $\zeta_q$ exponents computed from a wind tunnel 
experiment \cite{ruiz95} are also shown. It can be seen that
Model A gives scaling exponents in agreement with experiments
(within the experimental error) for $p \gtrsim 0.9$.

\subsection{Model B}
\begin{figure}
\includegraphics[scale=0.8]{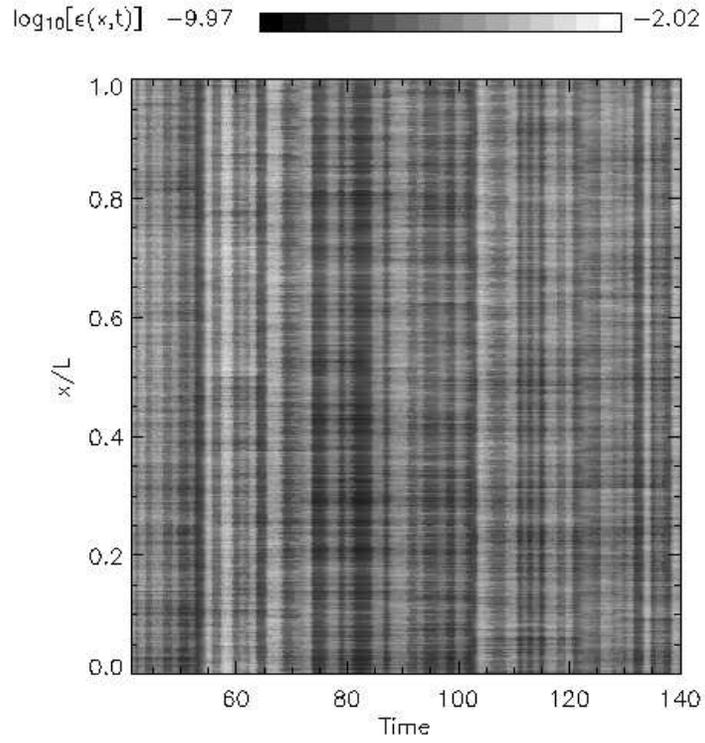}
\includegraphics[scale=0.8]{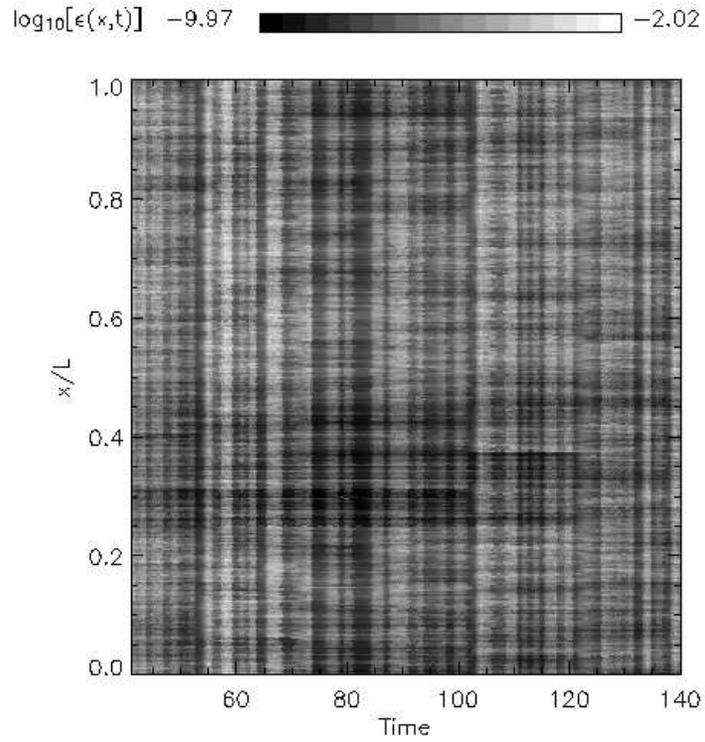}
\caption{Space-time structure of the energy dissipation rate $\varepsilon(x,t)$
in Model B for $p=0.7$ (top panel) and for $p=0.8$ (bottom panel). The grey 
levels refer to the logarithm of $\varepsilon(x,t)$.}
\label{fig-diss-b}
\end{figure}

The space-time structure of the energy dissipation rate $\varepsilon(x,t)$
for $p=0.7$ and $p=0.8$ is shown in Fig. \ref{fig-diss-b}. A larger
spatial fragmentation of $\varepsilon(x,t)$ for the larger $p$ is clearly
observed.

\begin{figure}
\includegraphics[width=\columnwidth]{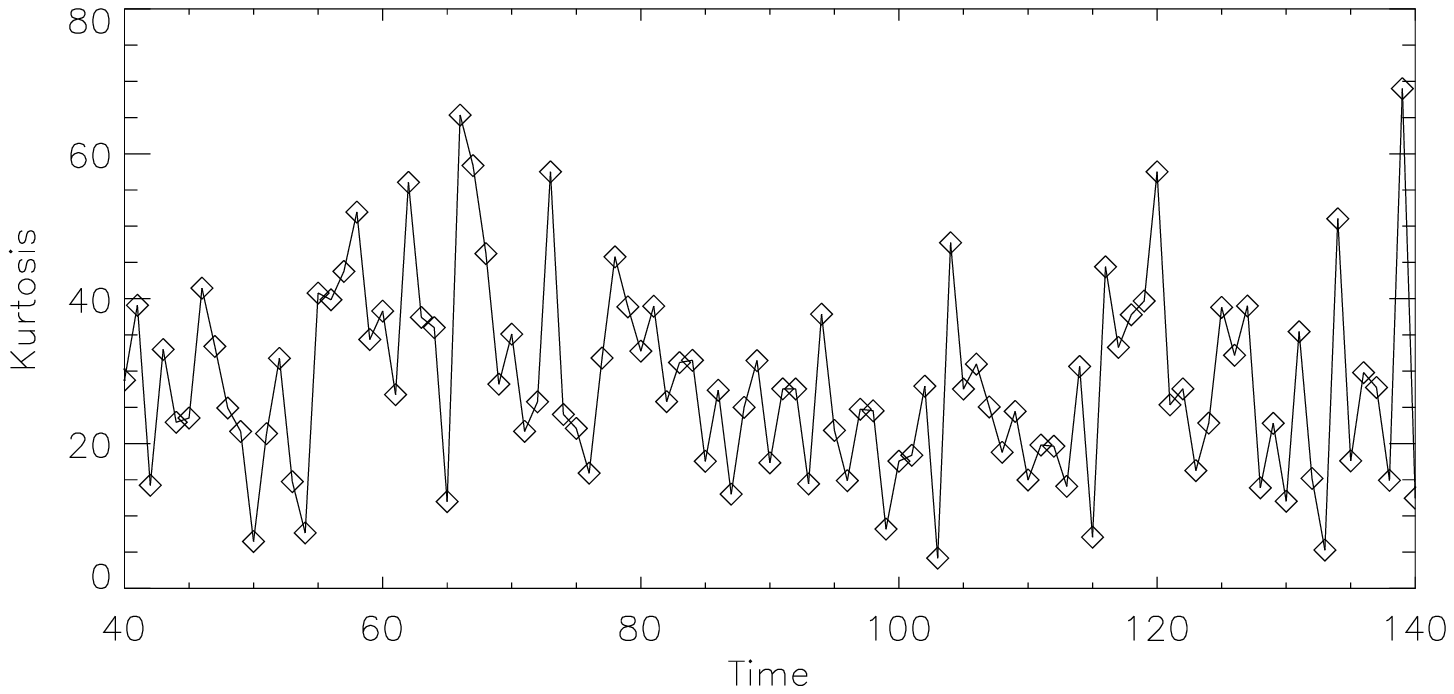}
\includegraphics[width=\columnwidth]{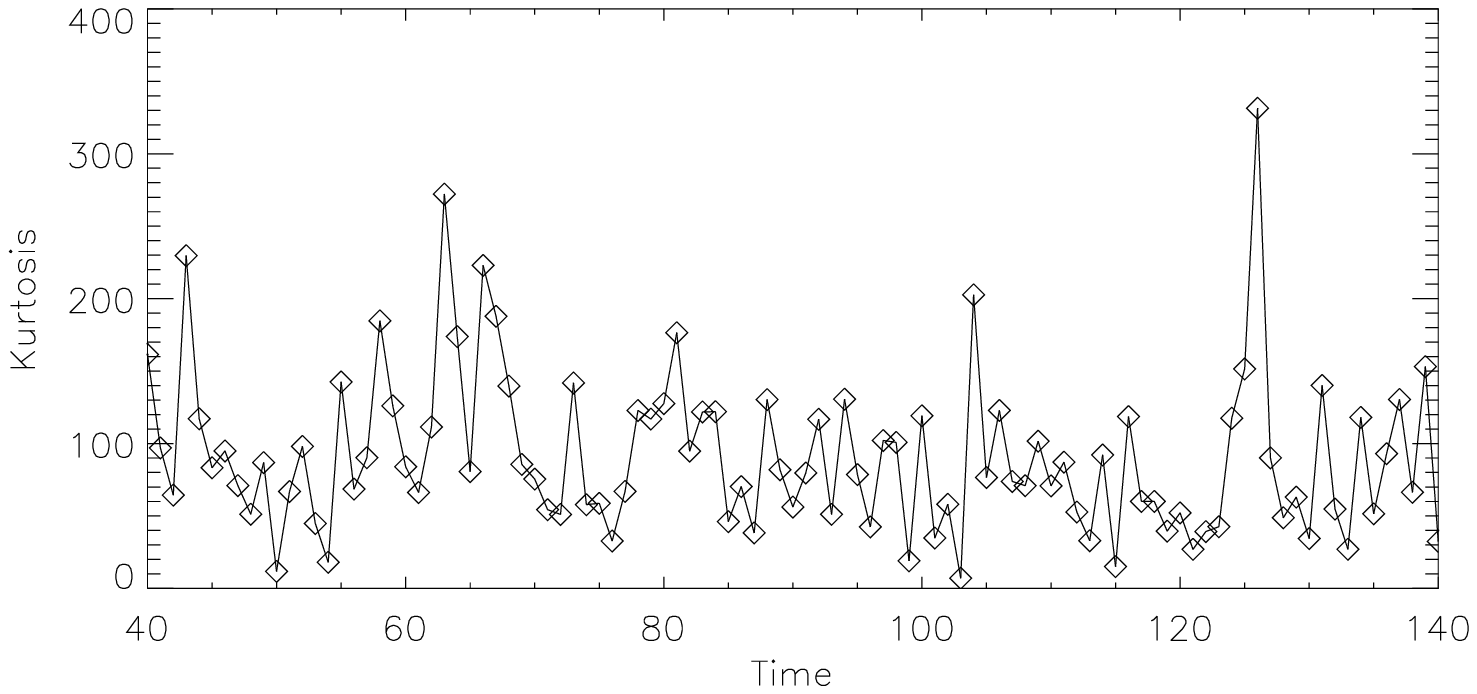}
\caption{Time evolution of the kurtosis of the spatial energy 
dissipation rate $\varepsilon(x)$ in Model B for $p=0.7$ (top panel) 
and $p=0.8$ (bottom panel).}
\label{fig-kurto-b}
\end{figure}

\begin{figure}
\includegraphics[width=\columnwidth]{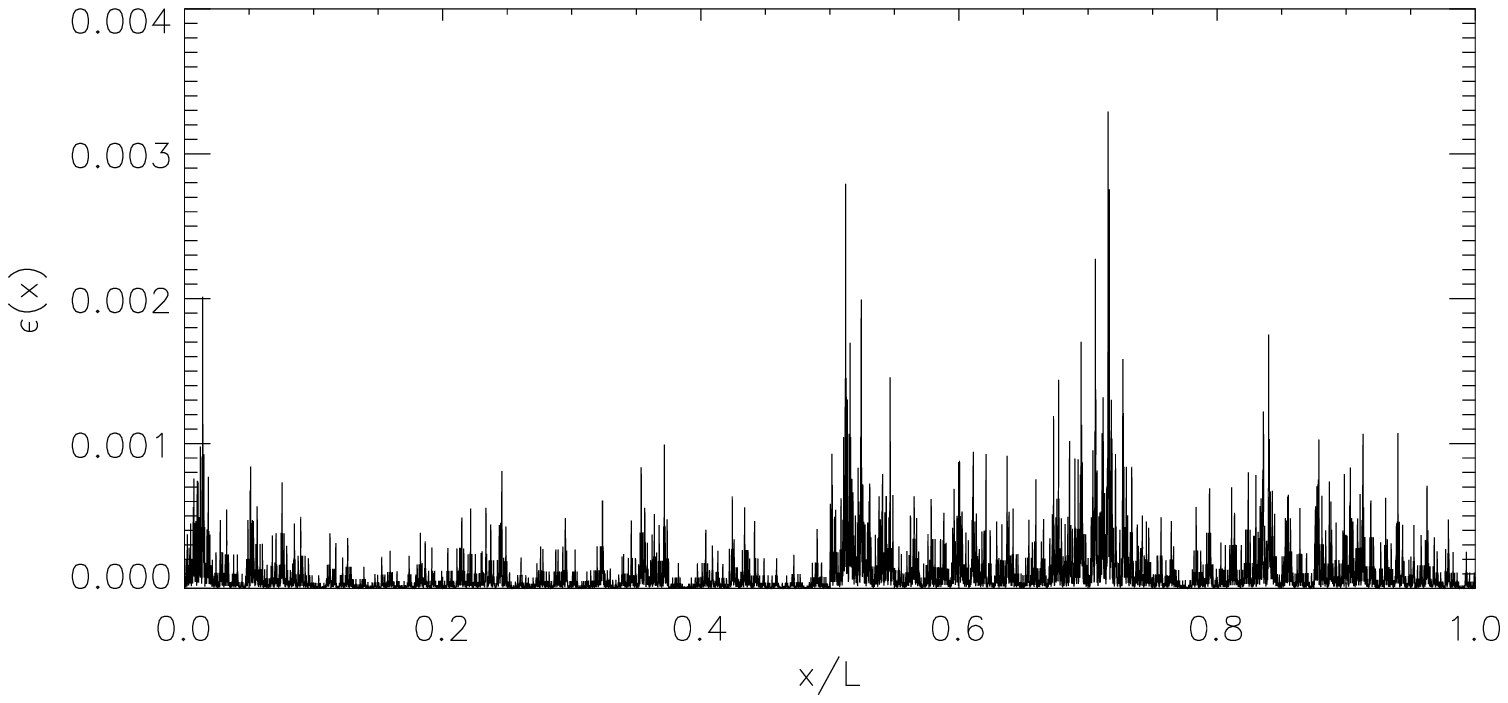}
\includegraphics[width=\columnwidth]{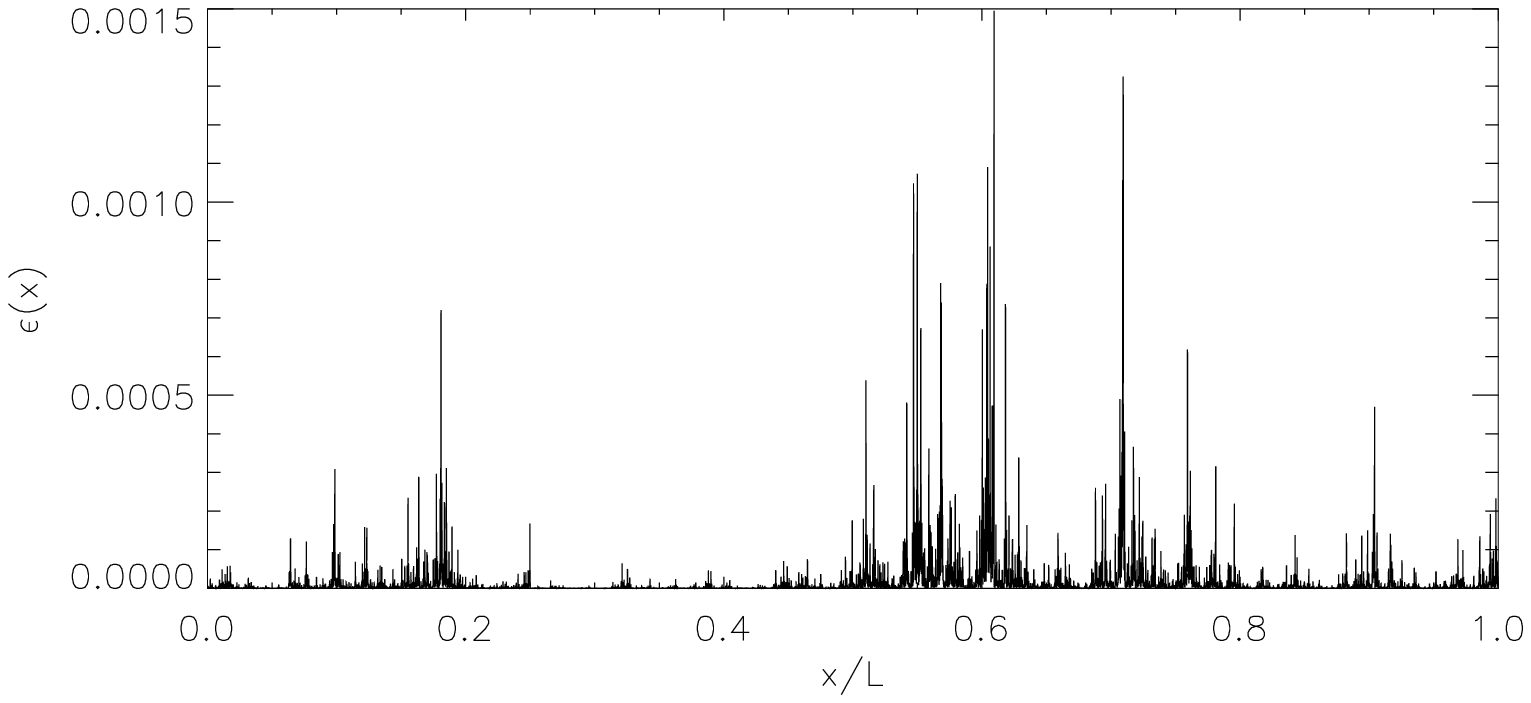}
\caption{Energy dissipation rate on the spatial domain in Model B for
$p=0.7$ and $t=58$ (top panel), $p=0.8$ and $t=66$ (bottom panel).}
\label{fig-ediss-b}
\end{figure}

The time evolution of the
kurtosis of $\varepsilon(x)$ for $p=0.7$ and $p=0.8$
is shown in Fig. \ref{fig-kurto-b}.
It can be seen that  the typical values of the kurtosis
increase going from $p=0.7$ to $p=0.8$ as it could be expected.
Comparing Fig. \ref{fig-kurto-b} to Fig. \ref{fig-kurto} we can notice
that, for the same $p$, Model B is much more intermittent
than Model A. For instance, using $p=0.8$  Model B
roughly reaches the same level of intermittency as Model A
with $p=1$. 
In Fig. \ref{fig-ediss-b} we show
the spatial structure of the energy dissipation rate $\varepsilon(x)$
at two time instants where a peak in the kurtosis is found, that is, 
$t=58$ for $p=0.7$ and $t=66$  for $p=0.8$.
From this figure it is clear that also for Model B 
$\varepsilon(x)$ shows very strong intermittency bursts
in space.

\begin{figure}
\includegraphics[width=\columnwidth]{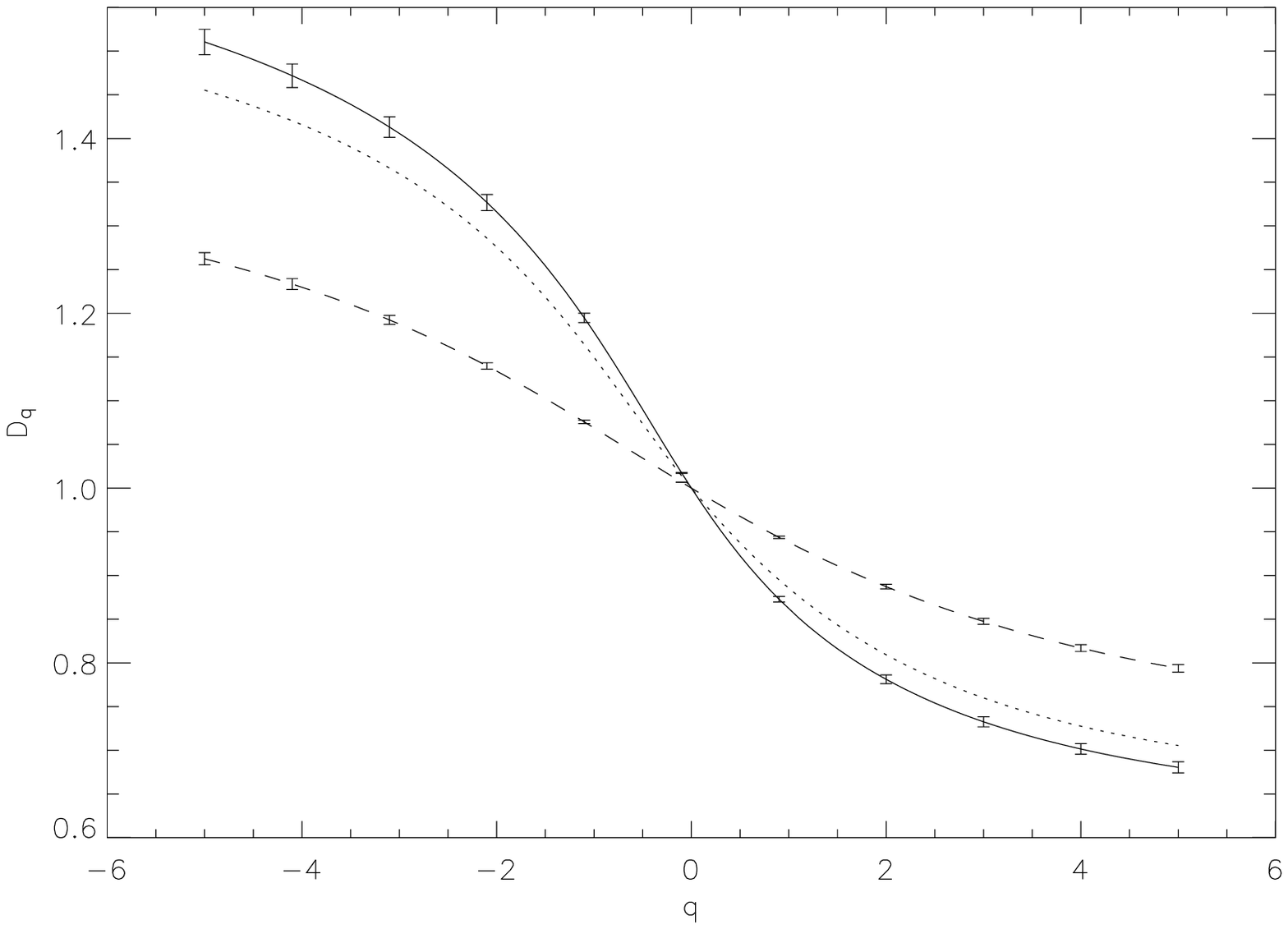}
\caption{Generalized dimensions $D_q$ of the spatial energy 
dissipation rates in Model B for $p=0.7$ ({\em Dashed line})
and  $p=0.8$ ({\em Solid line}). The $D_q$ values were calculated
as time averages over all the time instants considered (see text).
Error bars, representing standard
deviation errors, are also reported for some values of $q$.
The $D_q$ curve obtained in Model A for $p=1$ is also shown
as a comparison (dotted line).}
\label{fig-dq-b}
\end{figure}

Also for Model B we investigated the spatial intermittency 
properties of the energy dissipation rate through the multifractal
analysis described previously. The moments $\langle \mu^q \rangle$ given in
Eq. (\ref{eq-moments}) were calculated also in this case
for $-5 \le q \le 5$ at the 100 time instants $t=41,42,...,140$.
Good scalings were found for all the $q$ values and the dimensions
$D_q$ were obtained as averages over all the time instants
considered.
Fig. \ref{fig-dq-b} shows the plot of $D_q$ vs. $q$ for $p=0.7$
and $p=0.8$.
As expected, the $D_q$ spectrum is wider for $p=0.8$. Moreover
we can observe that the $D_q$ curve for $p=0.8$ is very close
to the one obtained in Model A for $p=1$.

The scaling exponents $\zeta_q$ of the velocity structure functions
in Model B as estimated from Eq. (\ref{eq-zetaq-dq}) are reported in
Table \ref{tab-vscal-b}. 
  \begin{table}
  \caption{Scaling exponents $\zeta_q$ for the velocity
  structure functions as 
  estimated from the energy dissipation scalings in Model B
  for $p=0.7$ and $p=0.8$. In the last column we report
  the velocity structure functions exponents computed from a wind tunnel 
  experiment \cite{ruiz95}.}   
  \begin{center} 
  \begin{tabular}{cccccc} 
  \hline
  $q$ \hspace{0.7cm} & $p$=0.7 \hspace{0.7cm} & $p$=0.8 \hspace{0.7cm}  
  & wind tunnel\\ 
  \hline
  1 \hspace{0.7cm} & $0.348 \pm 0.004$ \hspace{0.7cm} & 
  $0.369 \pm 0.009$ \hspace{0.7cm} & $0.37 \pm 0.01$ \\
  2 \hspace{0.7cm} & $0.681 \pm 0.004$ \hspace{0.7cm} & 
  $0.700 \pm 0.008$ \hspace{0.7cm} & $0.70 \pm 0.01$ \\
  3 \hspace{0.7cm} & $1.00$ \hspace{0.7cm} & $1.00$ \hspace{0.7cm}  
  & $1.00$ \\
  4 \hspace{0.7cm} & $1.307 \pm 0.006$ \hspace{0.7cm} &
  $1.277 \pm 0.014$ \hspace{0.7cm} & $1.28 \pm 0.02$ \\
  5 \hspace{0.7cm} & $1.602 \pm 0.015$ \hspace{0.7cm} &
  $1.536 \pm 0.031$ \hspace{0.7cm} & $1.54 \pm 0.03$ \\
  \hline
  \end{tabular} 
  \label{tab-vscal-b} 
  \end{center} 
  \end{table}
%
A good agreement with the scaling exponents
found in the wind tunnel experiment analyzed in Ref. \cite{ruiz95}
is obtained for $p=0.8$.

The fact that for a given $p$ a larger intermittency is found in
Model B than in Model A is clearly a consequence of the fact that
in the procedure used in Model B for the time evolution of the spatial 
distribution of $p$ and $1-p$ we consider also the local dynamics
of the energy cascade as pointed out in Section \ref{sec-method}.


\section{Conclusions}
\label{sec-conclu}

A problem which often arise when studying turbulent phenomena
occurring in astrophysical and space fluids is the description
of the intermittency of the energy dissipation process
from a spatio-temporal point of view. Due to the huge Reynolds
numbers occurring typically in these situations a dynamical
system modeling of space-time intermittency can represent an
important ingredient for the characterization of such systems.

In this paper we propose a method to model the main intermittency
features of energy dissipation in a turbulent system both in space
and time. This is done by using a turbulence shell model and
introducing some heuristic rules, partly inspired by the well
known cascade $p$-model, to construct a spatial structure
for the energy dissipation rate. 

To the aim of validating the model, we performed a series
of numerical simulations to study the spatial intermittency properties
of the energy dissipation rate for different values of the free parameter $p$.
The results show that the level
of spatial intermittency of the system can be simply tuned,
in both the proposed versions of the model,
by changing the value of $p$. The spatial intermittency of the system
is enhanced by increasing the $p$ parameter. Scaling laws in agreement with those
obtained in experiments involving fully turbulent hydrodynamic flows
are recovered in Model A for $0.9 \lesssim p \lesssim 1$, and in Model
B for $p \simeq 0.8$.

The results of this work open the way to applications of the proposed
model to different physical situations. In our opinion this model
could represent a useful tool to simulate the spatio-temporal
intermittency of turbulent energy dissipation in those high Reynolds
number astrophysical fluids where impulsive energy release processes
can be associated to the dynamics of the turbulent cascade.
To give just some examples, such a modeling could be interesting 
for studying the role of intermittent energy dissipation in the
active regions of the solar corona, in the interstellar medium, and
in accretion disks. We plan to investigate some of these problems in
future studies.

\begin{acknowledgments}
This work was partially supported by European Commission under contract MERG-6-CT-2005-014587.
\end{acknowledgments}


\end{document}